\documentclass[onecolumn,fleqn,10pt]{wlscirep}
\usepackage{dcolumn}
\usepackage{bm}
\usepackage{epsfig}
\usepackage{graphicx}
\usepackage{epstopdf}
\usepackage{amssymb}
\usepackage{amsfonts}
\usepackage{hyperref}
\usepackage{color,soul}
\usepackage{appendix}
\usepackage{wrapfig}

\title{Chiral topological excitons in the monolayer transition metal dichalcogenides}
\author[1,*]{Z. R. Gong}
\author[1]{W. Z. Luo}
\author[1]{Z. F. Jiang}
\author[1]{H. C. Fu}
\affil[1]{College of Physics and Energy, Shenzhen University, Shenzhen, 518060, P. R. China}

\affil[*]{gongzr@szu.edu.cn}

\begin{abstract}
We theoretically investigate the chiral topological excitons emerging in the monolayer transition metal dichalcogenides, where a bulk energy gap of valley excitons is opened up by a position dependent external magnetic field. We find two emerging chiral topological nontrivial excitons states, which exactly connects to the bulk topological properties, i.e., Chern number =2. The dependences of the spectrum of the chiral topological excitons on the width of the magnetic field domain wall as well as the magnetic filed strength are numerically revealed. The chiral topological valley excitons are not only important to the excitonic transport due to prevention of the backscattering, but also give rise to the quantum coherent control in the optoelectronic applications.
\end{abstract}
\begin{document}

\maketitle
\flushbottom
\maketitle
\thispagestyle{empty}

\section*{Introduction}

The topological states on
the boundary or surface is one of the most fascinating phenomena in
solid state physics. The electronic topological states have been extensively
investigated both theoretically~\cite{Hasan10,Qi10,Moore10,Qi11,Fu07}
and experimentally~\cite{Bernevig06,Konig07,Hsieh08,Xia09,Zhang09}
based on the realization of the spin-orbit interaction in topological
insulator. The topology protected conducting edge states have distinct
properties from bulk states, and
play important role in the electronic
transport and
facilitate the implementation of the topological states
based electronic devices. The photonic topological states are also
found
in various systems such as microwave range photonic crystals~\cite{Wang09,Poo11},
the arrays of coupled optical resonators and waveguides~\cite{Cho08,Koch10,Umucalilar11,Hafezi11,Hafezi13,Liang13,Jia13}
and metamaterials~\cite{Yannopapas12,Khanikaev13,Chen14}. It is motivated
by the same idea of the electronic counterpart that the chiral edge
states are prevent from backscattering and thus insensitive to disorder.
Recently this idea has been generalized to the system consisting
composite particles such as excitons and polarons~\cite{Karzig15,Bardyn15,Nalitov15}.

The topological states on the boundary of the 2D materials have attracted
a lot of interests due to its 1D nature of the boundary~\cite{Ryu02,Li12}.
It is the band inversion at the Dirac points
that leads to the band structure of the topological states in 2D
materials, such as graphene~\cite{Bery06,Peres06,Neto09,Castro08,Yao09,Qiao12}.
As a new member of the 2D materials family, the monolayer TMDs,
realized in laboratories recently, shed light on the valleytronics, exciton
physics and photoelectronic applications. The monolayer TMDs are direct
bandgap semiconductors, where the conduction and valence band edges
locate at the doubly degenerate corners of the Brillouin zone, as
known as the Dirac points~\cite{Xiao12,Xu14,Liu15}. In fact there
are two obstacles to generate electronic topological states in monolayer
TMDs through the same mechanism in graphene: one is that as the direct
gap semiconductor the TMDs possesses huge band gap ($\sim1.6-2$eV);
and another is that both the valence and conduction bands consist
of the transition metal $d$ orbitals on the same site, which prevents the staggered sublattice potential induced band inversion in graphene~\cite{Yao09}. Those properties basically forbid the electronic band inversion in pristine TMDs.

Nevertheless, we can study the realization of the composite particles
such as valley excitons instead of the electrons in the TMDs. For
the bright excitons, there are two valley pseudospin configurations
where the electron and hole both locate at either the $\mathrm{K}$
or $-\mathrm{K}$ valley~\cite{Xiao12,Xu14,Liu15}. The valley excitons
follow optical selection rule, which means that the valley exciton locating
at $\mathrm{K}$($-\mathrm{K}$) valley only couples to $\sigma^{+}$($\sigma^{-}$)
circularly polarized light field~\cite{Mak12,Zeng12,Cao12,Jones13,Wang14,Kadantsev12,Niu15}.
Additionally, the Coulomb interaction between the electron and hole
is exceptionally strong because of the 2D quantum confinement. The
TMDs possesses valley degree of freedom, the valley-related optical
selection rule and the strong Coulomb interaction, offering a new
2D system to explore the exciton physics. In addition, the two kinds
of the valley excitons have opposite responses to the external out-of-plane
magnetic field~\cite{Aivazian15}, which implies a possible band inversion
for valley excitons. In this sense, the TMDs provide a unprecedent
platform to investigate the topological states of valley excitons.

In this paper, we shall theoretically investigate the chiral topological
excitons emerging in the monolayer TMDs, where a bulk energy
gap of valley excitons is opened up by a position dependent external
magnetic field. The band dispersion of the bulk valley exciton realizes
the massive Dirac cone, which exploits the strong valley-orbital coupling induced by Coulomb exchange
interaction and the valley Zeeman splitting in the external magnetic
field. For this unique Dirac cone of valley excitons, there are chiral
topological exciton states emerging in the gap, whose number is determined
by the bulk topological properties, i.e., Chern number =2. Since the
time reversal symmetry is broken by the magnetic field, there are
only two different chiral topological exciton modes. We numerically
reveal the dependences of the spectrum of the chiral topological excitons
on the width of the magnetic field domain wall as well as the magnetic
filed strength. The chiral topological valley excitons are not only
important to the excitonic transport due to prevention of the backscattering,
but also have potential application to the quantum coherent control in the optoelectronics.

\section*{Results}
\subsection*{The bulk topological properties}

The precise state of the valley exciton at $\mathrm{K}$ point
is the superposition of the electron-hole
pairs with all possible wave vector of the relative motion $\mathbf{q}$
and the definite wave vector of the center-of-mass motion $\mathbf{k}$
\begin{eqnarray}
\left|\mathrm{+}\right\rangle _{\mathbf{k}}
& = & \exp(-i\mathbf{k}\mathbf{\cdot r})
\sum_{\mathbf{q}}\phi\left(\mathbf{q}\right)
e_{\mathbf{q+\frac{k}{2}},\uparrow}^{\dagger}
h_{\mathbf{-q+\frac{k}{2},}\Downarrow}^{\dagger}
\left|\mbox{vac}\right\rangle,
\label{eq1}
\end{eqnarray}
with the profile of relative motion $\phi\left(\mathbf{q}\right)$,
the electron (hole) creation operators $e^{\dagger}$($h^{\dagger}$)
and the electron (hole) spin up $\uparrow$ (spin down$\Downarrow$).
The quantum state $\left|-\right\rangle _{\mathbf{k}}$
of the another valley exciton at $-\mathrm{K}$ point
is the time reversal of the
above one.
In the low excitation limit of the valley excitons, the
Hilbert space is spanned by the pseudospins described by the spin up(down)
state $\left|+\right\rangle _{\mathbf{k}}$
($\left|-\right\rangle _{\mathbf{k}}$). We start from the Hamiltonian
of the valley excitons in the out-of-plane magnetic field
\begin{eqnarray}
H\left(\mathbf{k}\right) & = & \sum_{\mathbf{k}}\left(H_{\mathrm{k}}+H_{\mathrm{E}}+H_{\mathrm{B}}\right),\label{eq2}
\end{eqnarray}
where $H_{\mathrm{k}}=\frac{\hbar^{2}k^{2}}{2M}I$ is the
kinetic energy of the valley excitons,
$H_{\mathrm{E}}=\overrightarrow{h}\left(k\right)\cdot\overrightarrow{\mathbf{\sigma}}+\left| \overrightarrow{h}\left(k\right)\right| \mathbf{I}$
is the valley-orbit coupling term with $\overrightarrow{\mathbf{\sigma}}=\left(\sigma_{x},\sigma_{y}\right)$,
and $H_{\mathrm{B}}=-g_{\mathrm{B}}\mu_{\mathrm{B}}B\left(\mathbf{r}\right)\sigma_{z}$ is the position dependent valley Zeeman splitting.
Here, $M$ is total effective mass of the electron and hole consisting
valley excitons, $\mathbf{k}=(k_{x},k_{y})=(k\cos\theta,k\sin\theta)$
and $\mathbf{r}=(x,y)$ are the wave vector and the position
coordinate of the valley excitons' center-of-mass motion respectively, $\mu_{\mathrm{B}}$
is the Bohr magneton, $I$ and $\sigma_{\alpha}\left(\alpha=x,y,z\right)$
are respectively the identity matrix and Pauli matrices, and $B\left(\mathbf{r}\right)$
is the position dependent magnetic field.

The valley-orbit coupling actually introduces the inter-valley transition,
where the effective field $\overrightarrow{h}\left(k\right)=-J\left(k\right)\left(\cos2\theta,\sin2\theta\right)$
originates from the long range part of the Coulomb exchange interaction\cite{Yu14},
where $J(k)=Jk/K$ is the valley orbit coupling strength scaling linearly
of the wave numbers $k$, and $K=4\pi/3a$ is the wave vector from the
$\mathrm{K}$ to $\Gamma$ point of the Brillouin zone. The constant
$J\approx2\pi\frac{Ka^{2}t^{2}}{\epsilon a_{B}^{2}\Delta^{2}}$ only
depends on the parameters of the monolayer TMDs, the lattice
constant $a$, the hooping constant $t$, the effective dielectric
constant $\epsilon$, the Bohr radius of the exciton $a_{B}$ and
the bang gap $\Delta.$

\begin{figure}[ht]
\centering
\includegraphics[viewport=26 485 564 790,clip=true,width=\linewidth]{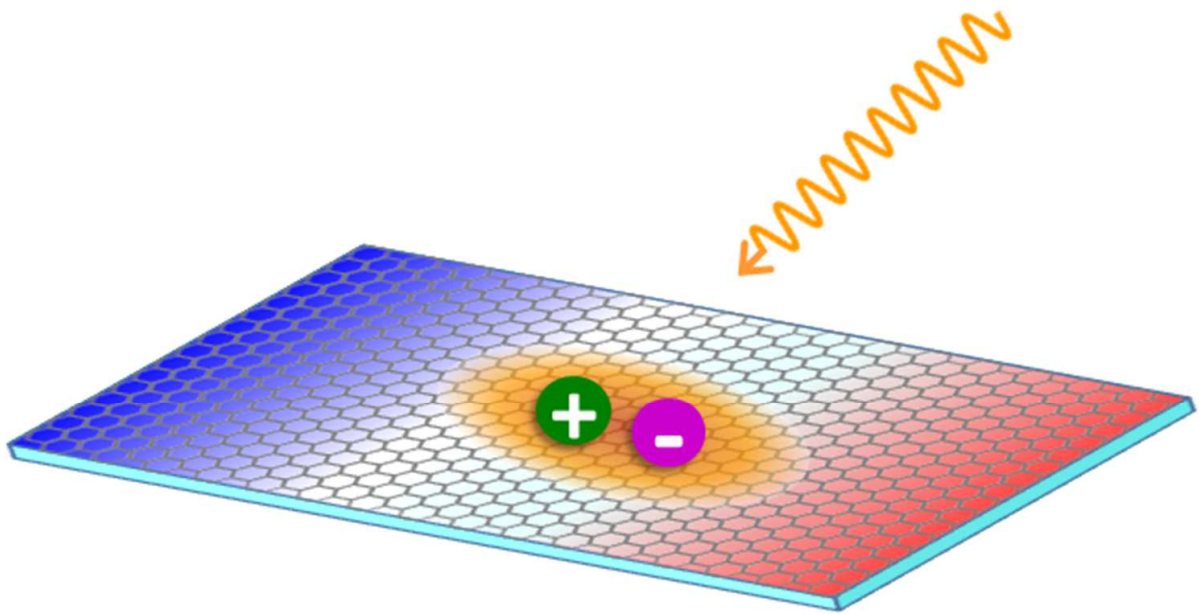}
\caption{Carton of generating the topological exciton in the
vicinity of the magnetic field domain wall. The orange wave line represents the pumping light field. The wavefuntion of its relative motion is
represented by the orange area, in which the electron and the hole are respectively represented by the purple sphere with "-" symbol and the green sphere with "+" symbol. The colored region of the monolayer TMDs represent the position dependent
magnetic field.}
\label{fig:fig1}
\end{figure}

The magnetic response of the valley excitons, caused by the out-of-plane
magnetic field, gives rise to the last term $H_{\mathrm{B}}$. Since
the valley exciton is roughly regarded as a bounded electron-hole
pair, both the electron and hole magnetic moments contribute to
the exciton's Zeeman splitting in the magnetic field, leading to the
effective g-factor of the valley excitons
\begin{eqnarray}
g_{\mathrm{B}}\left(\mathbf{k}\right) & = & \sum_{\mathbf{q}}\left|\phi\left(\mathbf{q}\right)\right|^{2}\left[g_{e}\left(\mathbf{q},\mathbf{k}\right)-g_{h}\left(\mathbf{q},\mathbf{k}\right)\right].\label{eq3}
\end{eqnarray}
The minus sign before the hole's g factor originates from the electron-hole
duality in the semiconductor. Both the transition metal $d$ orbital
and valley magnetic moments contribute to the Lande g-factor of electron
and hole. In the recent experiment based on $\textrm{WSe}_2$, the typical measured $g_B$ is about
1.8~\cite{Aivazian15}. In this sense, the typical valley Zeeman splitting is about several meV when the applied magnetic field is up to $10T$. Since the momentum scale of the profile
$\phi\left(\mathbf{q}\right)$ is about the reciprocal of the of Bohr
radius of the valley exciton, which is much smaller than the momentum
scale of the Lande g-factor of electron and hole. In this sense, we
obtain the approximate effective Lande g-factor for the valley excitons
$g_{\mathrm{B}}\approx g_{\mathrm{B}}\left(\mathbf{k}=0\right)$,
which becomes independent of the wave vector of the center-of-mass
motion. The $\sigma_{z}$ term in the $H_{\mathrm{B}}$ implies that the valley
Zeeman energies are exactly opposite at $\mathrm{K}$ and $-\mathrm{K}$
points, which results from the valley-index-dependent magnetic moments
of the valley excitons. The setup is schematically shown in Fig.~\ref{fig:fig1}.
It is noticed that the inhomogeneity of the magnetic field $U(r_\alpha)  (\alpha=e,h)$ shifts the momentum of the electron and hole and eventually cause the overlap of their wavefunctions, which gives rise to an additional inter-valley coupling of single states. It is straightforwardly calculated as

\begin{eqnarray}
V_{inter}^{\alpha} & = & \left\langle \alpha_{\mathbf{K+p}}\right|U\left(\mathbf{r_{\alpha}}\right)\left|\alpha_{-\mathbf{K+q}}\right\rangle \nonumber \\
 & = & \int d\mathbf{r_{\alpha}}u_{\mathbf{K+p}}^{*}\left(r_{\alpha}\right)\exp\left(-i\mathbf{\left(K+p\right)\cdot r_{\alpha}}\right)U\left(\mathbf{r_{\alpha}}\right)\times\nonumber \\
 &  & u_{-\mathbf{K+q}}\left(r_{\alpha}\right)\exp\left(i\mathbf{\left(-K+q\right)\cdot r_{\alpha}}\right)\nonumber \\
  & \approx & \frac{1}{V}d\mathbf{r_{\alpha}}U\left(\mathbf{r_{\alpha}}\right)\exp\left(-2i\mathbf{K\cdot\mathbf{r_{\alpha}}}\right),
\end{eqnarray}
where $\left|\alpha_{\mathbf{k}}\right\rangle =\exp\left(i\mathbf{k}\cdot\mathbf{\ensuremath{r}}_{\alpha}\right)\left|\mathbf{u_{k}\left(r_{\alpha}\right)}\right\rangle$ are the Bloch wave functions of the electron and hole. In the last step we apply the condition $\mathbf{p},\mathbf{q} \ll \mathbf{K}$. For a slowly varying magnetic field domain wall with a typical length of domain wall more than hundreds of lattice constant, the inter-valley couplings of the single states are about $10^{-5} E_B$, where $E_B$ is the Zeeman energy. Obviously, they are sufficiently small. Additionally, since the inter-valley coupling of the excitons basically are the summation of the inter-valley coupling of single states, both of them are neglected in the following discussion.

In the bulk, the Hamiltonian can be written as a matrix form~\cite{Yu15}
\begin{eqnarray}
H_{\mathrm{bulk}} & = & \sum_{\mathbf{k}}\left(\frac{\hbar^{2}k^{2}}{2M}+J\left(k\right)\right)I+\left[\begin{array}{cc}
\Delta & -J(k)e^{-2i\theta}\\
-J(k)e^{2i\theta} & -\Delta
\end{array}\right]\label{eq4}
\end{eqnarray}
on the basis
$ \{ \left|\mathrm{+}\right\rangle_{\mathbf{k}},
     \left|\mathrm{-}\right\rangle_{\mathbf{k}} \}$,
where
$\Delta=g_{\mathrm{B}}\mu_{\mathrm{B}}B$ is the valley Zeeman energy.
The valley Zeeman energy approximately takes the fixed value because
we only consider the bulk topological properties at the region
where the applied magnetic field is homogeneous.
The valley exciton dispersion splits into two branches with energies
\begin{eqnarray}
E_{\pm}\left(\mathbf{k}\right) & = & \frac{\hbar^{2}k^{2}}{2M}+J\left(k\right)\pm\sqrt{\Delta^{2}+J(k)^{2}}\label{eq5}
\end{eqnarray}
and corresponding eigen-wavefunctions
\begin{eqnarray}
\left|u_{+}\left(\mathbf{k}\right)\right\rangle  & = & \cos\frac{\alpha}{2}\left|+\right\rangle _{\mathbf{k}}-\sin\frac{\alpha}{2}e^{2i\theta}\left|-\right\rangle _{\mathbf{k}},\\
\left|u_{-}\left(\mathbf{k}\right)\right\rangle  & = & \sin\frac{\alpha}{2}e^{-2i\theta}\left|+\right\rangle _{\mathbf{k}}+\cos\frac{\alpha}{2}\left|-\right\rangle _{\mathbf{k}},
\end{eqnarray}
where $\tan\alpha=\frac{J\left(k\right)}{\Delta}$.
So it realizes a massive Dirac cone. The valley Zeeman energy plays
the role of the mass in the Dirac-like equation and opens up a gap
between two bands of valley excitons (see Fig.2(a)). As composite
particles, the valley excitons still share the same Brillouin zone
of the electron and hole in the monolayer TMDs. In contrast of the
Dirac cones of the electron which locates at the corner of the Brillouin
zone, the unique Dirac cone of the valley exciton locates at the center
of the Brillouin zone.

In order to describe the bulk topological property of the valley excitons,
the Berry connection $\mathcal{A}\left(\mathbf{k}\right)$ and
Berry curvature $\Omega\left(\mathbf{k}\right)$, defined respectively as
\begin{eqnarray}
\mathcal{A}\left(\mathbf{k}\right) & \equiv & \left\langle u_{-}\left(\mathbf{k}\right)\right|i\nabla_{\mathbf{k}}\left|u_{-}\left(\mathbf{k}\right)\right\rangle \label{eq8} \\
\Omega\left(\mathbf{k}\right) & \equiv & \nabla_{\mathbf{k}}\times\mathcal{A\left(\mathbf{k}\right)}\label{eq9}
\end{eqnarray}
are introduced as the gauge potential and the gauge field of the lower
valley exciton band. The Berry curvature is regarded as a magnetic
field in the valley exciton center-of-mass momentum space, the integral
of which over the $\mathbf{k}$-space area gives rise to the Berry
phase of the valley exciton if it adiabatically go around the area
boundary. The Chern invariant is defined as the flux of the Berry
curvature threading the entire Brillouin zone
\begin{eqnarray}
\mathcal{C} & = & \frac{1}{2\pi} \int_{\mathrm{BZ}} d\mathbf{k} \Omega\left(\mathbf{k}\right).
\label{eq10}
\end{eqnarray}
For the valley excitons described by equation (2), one find the Berry curvature
centered at the Dirac cone
\begin{eqnarray}
\Omega\left(\mathbf{k}\right) & = &
\frac{J^{2}\Delta K}{\left(J^{2}k^{2}+\Delta^{2} K^{2} \right)^{\frac{3}{2}}}
\label{eq11}
\end{eqnarray}
and thus the Chern invariant $\mathcal{C}={\rm sign}\left(\Delta\right)\text{.}$
The nonzero Chern number implies the existence of topology states
of valley excitons.

When the position dependent magnetic field is applied, it leads to
the band inversion of the valley excitons, and the number of the topological
charge equals to the difference of the bulk topological charges on
the both side of the domain wall~\cite{Yao09}, which reads $\nu=\mbox{\ensuremath{\mathcal{C}\left({\rm left}\right)}-\ensuremath{\mathcal{C}\left({\rm right}\right)}=2.}$
Therefore, it is imaginable that two topological states will emerge
at the vicinity of the magnetic field domain wall. Since the magnetic
field breaks the time reversal symmetry and the Dirac cone is uniquely
centered at zero momentum point, such topological states become chiral
ones without time reversal symmetry.

For the sake of simplicity, we assume
the position-dependent magnetic field varies only along $x$-direction,
namely $B\left(\mathbf{r}\right)\equiv B\left(x\right)$. When the magnetic
field varies slowly along $x$-direction, the wavefunction of the
topological exciton can be written as a two-component vector
\begin{eqnarray}
\Phi\left(\mathbf{r}\right) & = & \left[\begin{array}{c}
\Phi_{+}\left(x\right)\\
\Phi_{-}\left(x\right)
\end{array}\right]e^{ip_{y}y}\label{eq12}
\end{eqnarray}
where $\Phi_{\pm}\left(x\right)$ are the wavefunction profile, and
$p_{y}$ is the $y$-component momentum. The wavefunction profile
$\Psi\left(x\right)=\{\Phi_{+}\left(x\right),\Phi_{-}\left(x\right)\}^{\mathrm{T}}$
satisfies the following equation
\begin{eqnarray}
H\left(-i\nabla\right)\Psi\left(x\right) & = & E\Psi\left(x\right),\label{eq13}
\end{eqnarray}
where the momentum in the original Hamiltonian is substituted by the
operator in the real space as $\mathbf{k}\rightarrow-i\nabla$. According
to the symmetry analysis of the above equation, there are two solutions
$\Phi_{+}^{1}\left(x\right)=\Phi_{-}^{1}\left(-x\right)$
and $\Phi_{+}^{2}\left(x\right)=-\Phi_{-}^{2}\left(-x\right)$ corresponding
to two topological excitons. However, the chirality index of the valley
exciton Dirac cone equals to 2 in contrast with the well known Dirac
cone, and the off-diagonal element $\propto ke^{-2i\theta}$
does not possess a simple operator form in the real space.
So it is appropriate to solve the equations in the momentum space numerically.
The details is presented in the Method section.

\subsection*{Numerical results.}
In order to confirm the existence of two chiral topological excitons, we numerically
evaluate the energy spectrum and the corresponding wavefunctions from  equation~(\ref{eq13}).
Here the constant $J$ is chosen as $1$eV, and the lattice constant for MoS$_2$ is $3.49\AA$.
In Fig.2 (a-c), the magnetic field domain wall is
assumed as $B\left(x\right)=B_{0}\tanh(x/l)$.
The typical spectrum of the topological excitons are demonstrated in Fig.2 (a).
Obviously there are two different topological excitons, which are
consistent with the Chern number. When
$q_{y}$ tends to positive (negative) infinity, the dispersion of
both topological excitons is convergent to the edge of the conduction
(valence) band. The dependence of the spectrum on the magnetic field strengthes, widths and
types of the magnetic domain wall is depicted in Fig.2 (b-d). In
Fig.2.(b) and (c), the solid and dashed lines corresponds to the first
and the second solution of the equation of wavefunction profile. The
red, blue and black lines in Fig.2 (b) correspond to different magnetic
valley Zeeman energy $E_{B}\equiv g_{\mathrm{B}}\mu_{\mathrm{B}}B_{{\rm max}}=1{\rm meV}\text{, 10meV, 50meV}$
and $l=500a$ with $a$ the lattice constant. Fig. 2(c) corresponds
to different width of the magnetic domain wall $l=100a, 500a, 1000a$
and $E_{B}=10$meV, respectively. Obviously, the smaller the width
of magnetic domain wall is, or the stronger the magnetic field strength
is, the larger the energy difference between two topological excitons
becomes. It actually results from the stronger quantum confinement
of the magnetic domain wall, which suggests to adjust the spectrum
of the topological excitons through both the magnetic field strength
and the width of the magnetic domain wall. We compare the different
types of the magnetic domain wall and find their spectrum almost coincide,
which results from the similar behavior of those functions in the
vicinity of $x=0$ as $\tanh\left(x/l\right)\sim\sinh(x/l)\sim x/l.$
It implies that the topological excitons basically locates at the
vicinity of the magnetic domain wall.

We also present the corresponding wavefunctions in Fig. 3. Since the
four components of the two solutions are related to each
other as $\Phi_{+}^{1}\left(x;q_{y}\right)=\Phi_{-}^{1}\left(-x;q_{y}\right)$
and $\Phi_{+}^{2}\left(-x;-q_{y}\right)=-\Phi_{-}^{2}\left(x;-q_{y}\right)$,
we only present the contour plot of wavefunction profile $\Phi_{+}^{1}\left(x\right)$
and $\Phi_{+}^{2}\left(x\right)$ versus $x$ and $q_{y}$ in Fig.
3. Parameters in Fig. 3(a) and (b) are the same as ones in Fig. 2(b)
and (c). Obviously under stronger quantum confinement with smaller
width of the magnetic field domain wall or stronger magnetic field
strength, the topological valley excitons become more local in the
vicinity of the domain wall.

\section*{Discussion}
The optoelectronic properties of the topological excitons depend on
their optical dipole defined as $D_{i}\equiv\hat{e}\cdot\left\langle \Phi^{i}\mathbf{\left(\mathbf{r}\right)\left|p\right|}0\right\rangle $,
$i=1,2$,
where $\hat{e}$ is the polarization of pumping light field and $\mathbf{p}$
is the electric dipole moment. Since the relative and center-of-mass
motions are independent for the topological exciton, the optical dipole
can be factorized as $D_{i}=A_{i}\left(D_{+}+(-1)^{i+1}D_{-}\right),$
where $A_{i}=\left|\int dx\Phi_{+}^{i}\left(x\right)\right|=\left|\int dx\Phi_{-}^{i}\left(x\right)\right|$
are the integrals of the wavefunction profile and $D_{\pm}=\hat{e}\cdot\left\langle \pm\left|\mathbf{p}\right|0\right\rangle $
are the optical dipole for the valley excitons $\left|\pm\right\rangle _{\mathbf{k=0}}$
. It indicates that the first and second topological excitons exactly
inherit the optical selection rules from the linear combinations of
the valley excitons $\left|+\right\rangle \pm\left|-\right\rangle $,
which means that the two topological excitons can be initialized by utilizing
the linear-polarized pumping light fields along $x$- and $y$-direction,
respectively. Additionally, the optical dipole are adjustable by tuning
the magnetic field because $A_{i}$ depends on the widths of the magnetic
field domain wall and magnetic field strength. These two topological
excitons possess controllable gap($\sim $meV), optical initialization
and robust transportation protected by topology, and thus may shed light on the
quantum coherent optoelectronic devices based on these topological
excitons in TMDs.

The typical time to generate the valley excitons is much shorter than the the lifetime of the valley excitons, which is about tens of ps for monolayer TMDs. Additionally, the optical dipole of the topological chiral excitons can be adjusted by the applied magnetic field, which suggests a longer lifetime for the topological excitons. Therefore, the lifetime of the excitons would not affect the generation of the topological excitons. The longer lifetime allows more subtle control, which may facilitates the application of the
optoelectronics based on the topological excitons.

\section*{Methods}

The spectrum and the corresponding wave-functions of the chiral topological
excitons can be obtained by solving equation~(\ref{eq13}). Actually, the quadratic
term together with the linear dispersion from the massless Dirac equation
gives rise to energy minimum around $\frac{k}{K}=\frac{JM}{K^{2}\hbar^{2}}\sim0.08$
corresponding to the minimum energy $-\frac{MJ^{2}}{2K^{2}\hbar^{2}}\sim-0.04$eV,
which is much larger than the magnetic field induced Zeeman splitting.
Therefore with the position dependent magnetic field $B(\mathbf{r})$
and considering the emergent edge state around Dirac point, we drop the
quadratic terms and obtain the Hamiltonian of topological excitons
as
\begin{eqnarray}
H_{\mathrm{ex}} & = & \sum_{\mathbf{k}}\left[\begin{array}{cc}
J\left(k\right)+\Delta & -J(k)e^{-2i\theta}\\
-J(k)e^{2i\theta} & J\left(k\right)-\Delta
\end{array}\right].\label{eq14}
\end{eqnarray}
Additionally, we define parameters $K_{B}\left(x\right)=Kg_{E}\mu_{B}B\left(x\right)/J$
and $K_{E}=EK/J$. The Schr\"{o}dinger equation for the wave-function
profile $\ensuremath{\Phi_{\pm}\left(x\right)}$ becomes
\begin{equation}
\begin{cases}
\left(k-K_{B}\left(x\right)-K_{E}\right)\ensuremath{\Phi_{+}\left(x\right)}-ke^{-2i\theta}\ensuremath{\Phi_{-}\left(x\right)}=0,\\
-ke^{-2i\theta}\ensuremath{\Phi_{+}\left(x\right)}+\left(k+K_{B}\left(x\right)-K_{E}\right)\ensuremath{\Phi_{-}\left(x\right)}=0.
\end{cases}
\label{eq15}
\end{equation}

Although the momentum operator can be written as $ke^{i\theta}=k_{x}+ik_{y}=-i\partial_{x}+\partial_{y}$
in the real space, the operators $k$ and $ke^{-2i\theta}$ do not
possess a simple operator form. However, they are classical numbers
in the momentum space. Thus it is convenient to solve above eigen-equations
in the momentum space. By applying the Fourier transformation, the
equation~(\ref{eq15}) becomes
\begin{equation}
\begin{cases}
-\int dp_{x}K_{B}\left(k_{x}-p_{x}\right)\Psi_{+}\left(p_{x}\right)+\left(k-K_{E}\right)\Psi_{+}\left(k_{x}\right)-\frac{\left(k_{x}-iq_{y}\right)^{2}}{\sqrt{k_{x}^{2}+q_{y}^{2}}}\Psi_{-}\left(k_{x}\right)=0,\\
-\frac{\left(k_{x}+iq_{y}\right)^{2}}{\sqrt{k_{x}^{2}+q_{y}^{2}}}\Psi_{+}\left(k_{x}\right)+\int dp_{x}K_{B}\left(k_{x}-p_{x}\right)\Psi_{-}\left(p_{x}\right)+\left(k-K_{E}\right)\Psi_{-}\left(k_{x}\right)=0,
\end{cases}\label{eq16}
\end{equation}
The convolution terms means all the equations for different $k_{x}$
are coupled to each other. If we discretize the momentum $k_{x}$,
all the equations are linear and thus the eigenvalues $K_{E}$
and the corresponding wavefunctions can be obtained numerically.

\section*{Acknowledgements (not compulsory)}

This work is supported by NSFC Grants No. 11504241 and the Natural Science Foundation of SZU Grants No. 201551.

\section*{Author contributions statement}

Z.R.G. proposed the project and H. C. F. supervised the project. W.Z.L. carried out the study. Z. F. J. provide the numerical calculations. All authors analysed the results and co-wrote the paper.

\section*{Competing financial interests}
The authors declare no competing financial interests.

\begin{figure}[ht]
\centering
\includegraphics[viewport=15 276 578 764,clip=true,width=\linewidth]{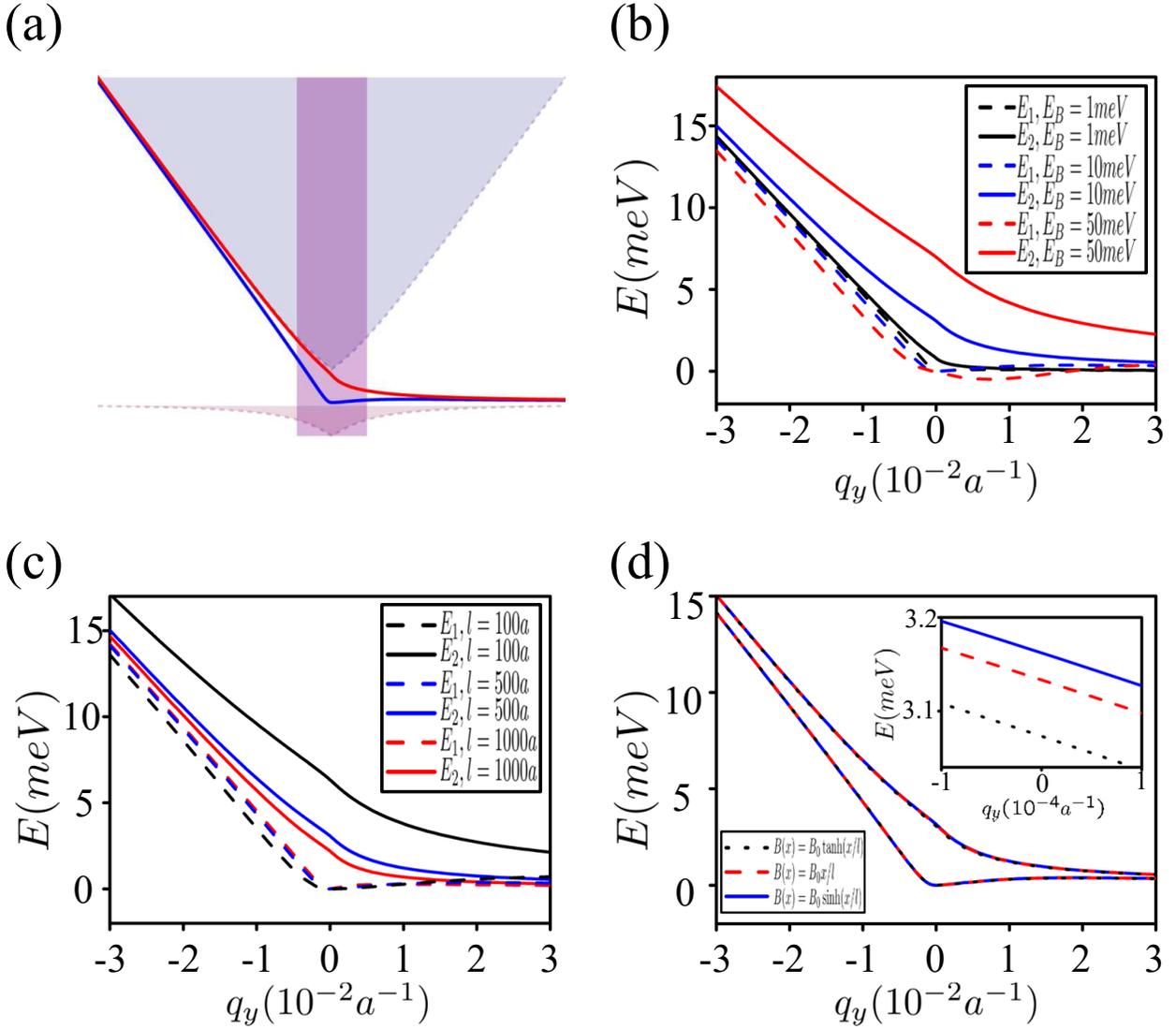}
\caption{(a) Typical spectrum of the topological excitons,
which are denoted by the red and blue solid lines. Here, the light
blue, light red and purple region respectively denotes the upper band,
the lower band and the light cone of the pumping light field. (b)
The spectrum of the topological excitons versus $q_{y}$ for different
magnetic field strength. (c) The spectrum of the topological valley
excitons versus $q_{y}$ for different widths of the magnetic domain
wall. (d) The spectrum of the topological excitons versus $q_{y}$
for different types of the magnetic domain wall. Insert: magnified
view of the spectrum. See text for the details.}
\label{fig:fig2}
\end{figure}

\begin{figure}[ht]
\centering
\includegraphics[viewport=15 101 578 781,clip=true,width=\linewidth]{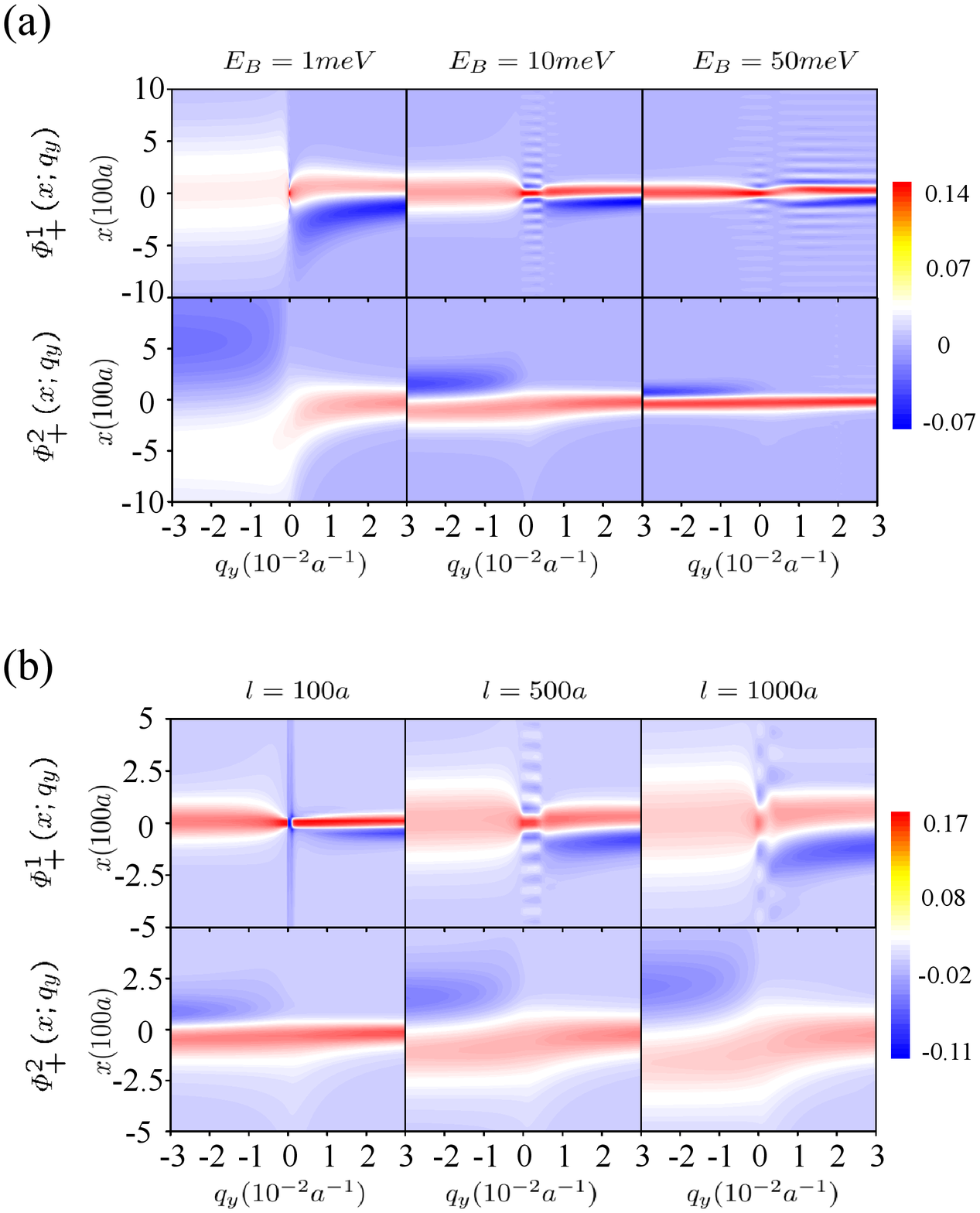}
\caption{Contour plot of the wavefunction profile $\Phi_{+}^{1}\left(x\right)$
and $\Phi_{+}^{2}\left(x\right)$ versus $x$ and $q_{y}$ for (a)
different widths of the magnetic field domain wall and (b) different
magnetic field strengths. Obviously under stronger quantum confinement
with smaller widths of the magnetic field domain wall or stronger
magnetic field strengths, the topological valley excitons becomes
more local in the vicinity of the domain wall.}
\label{fig:fig3}
\end{figure}


\begin{thebibliography}{99}
\bibitem{Hasan10} Hasan, M. Z. \& Kane, C. L. Colloquium: Topological
insulators. \textit{Rev. Mod. Phys.} \textbf{82}, 3045 (2010).

\bibitem{Qi10} Qi, X.-L. \& Zhang, S.-C. The quantum spin Hall effect
and topological insulators. \textit{Phys. Today}$\text{\textbf{63}}$,
33 (2010).

\bibitem{Moore10} Moore, J. E. The birth of topological insulators.
\textit{Nature}$\textbf{464}$,194 (2010).

\bibitem{Qi11} Qi, X.-L. \& Zhang S.-C. Topological insulators and
superconductors. \textit{Rev. Mod. Phys.} $\textbf{83}$, 1057 (2011).

\bibitem{Fu07} Fu, L. \& Kane, C. L. Topological insulators with
inversion symmetry. \textit{Phys. Rev. B}$\textbf{76}$, 045302 (2007).

\bibitem{Bernevig06} Bernevig, B. A., Hughes, T. L. \& Zhang S. C.,
Quantum Spin Hall Effect and Topological Phase Transition in HgTe
Quantum Wells. \textit{Science} $\textbf{314}$, 1757 (2006).

\bibitem{Konig07} K\"{o}nig, M. et al., Quantum Spin Hall
Insulator State in HgTe Quantum Wells. \textit{Science} $\textbf{318}$,
766 (2007).

\bibitem{Hsieh08} Hsieh, D. et al., A topological Dirac insulator in a
quantum spin Hall phase. \textit{Nature(London)} $\textbf{452}$,
970 (2008).

\bibitem{Xia09} Xia, Y. et al., Observation of a large-gap topological-insulator class with
a single Dirac cone on the surface. \textit{Nat. Phys.} $\textbf{5}$,
398 (2009).

\bibitem{Zhang09} Zhang, H. et al., Topological insulators in $\mathrm{Bi_{2}Se_{3}}$,
$\mathrm{Bi_{2}Te_{3}}$ and $\mathrm{Sb_{2}Te_{3}}$ with a single
Dirac cone on the surface. \textit{Nat. Phys.} $\textbf{5}$, 438
(2009).

\bibitem{Wang09} Wang, Z., Chong, Y., Joannopoulos, J. D. \& Solja\v{c}i\'{c},
M. Observation of unidirectional backscattering-immune topological
electromagnetic states. \textit{Nature} $\textbf{461}$, 772 (2009).

\bibitem{Poo11} Poo,Y. et al., Experimental Realization of Self-Guiding Unidirectional Electromagnetic
Edge States. \textit{Phys. Rev. Lett.} $\textbf{106}$, 093903 (2011).

\bibitem{Cho08} Cho, J., Angelakis, D. G. \& Bose, S. Fractional
Quantum Hall State in Coupled Cavities. \textit{Phys. Rev. Lett.}$\textbf{101}$,
246809 (2008).

\bibitem{Koch10} Koch, J., Houck, A. A., Hur, K. L. \& Girvin, S.
M. Time-reversal-symmetry breaking in circuit-QED-based photon lattices.
\textit{Phys. Rev. A} $\textbf{82}$, 043811 (2010).

\bibitem{Umucalilar11} Umucalilar, R. O. \& Carusotto, I. Artificial
gauge field for photons in coupled cavity arrays. \textit{Phys. Rev.
A} $\textbf{84}$, 043804 (2011).

\bibitem{Hafezi11} Hafezi, M., Demler, E. A., Lukin, M. D. \& Taylor,
J. M. Robust optical delay lines with topological protection. \textit{Nat.
Phys.} $\textbf{7}$, 907 (2011).

\bibitem{Hafezi13} Hafezi, M. et al., Imaging topological edge states in silicon photonics.
\textit{Nat. Photon.}$\textbf{7}$, 1001 (2013).

\bibitem{Liang13} Liang, G. Q. \& Chong, Y. D. Optical Resonator
Analog of a Two-Dimensional Topological Insulator. \textit{Phys. Rev.
Lett.} $\textbf{110}$, 203904 (2013).

\bibitem{Jia13} Ningyuan, J. et al., Time- and Site-Resolved Dynamics in a Topological Circuit.
\textit{Phys. Rev. X} $\textbf{5}$, 021031 (2015).

\bibitem{Yannopapas12} Yannopapas, V. Topological photonic bands
in two-dimensional networks of metamaterial elements. \textit{New
J. Phys.}$\textbf{14}$, 113017 (2012).

\bibitem{Khanikaev13} Khanikaev, A. B. et al., Photonic topological
insulators. \textit{Nat. Mater.}$\textbf{12}$, 233 (2013).

\bibitem{Chen14} Chen, W.-J. et al., Experimental realization of photonic
topological insulator in a uniaxial metacrystal waveguide. \textit{Nat.
Commum.}$\textbf{5}$, 5782 (2014).

\bibitem{Karzig15} Karzig, T., Bardyn, C.-E., Lindner, N. H. \& Refael,
G. Topological Polaritons. \textit{Phys. Rev. X} $\textbf{5}$, 031001
(2015).

\bibitem{Bardyn15} Bardyn, C.-E., Karzig, T., Refael, G. \& Liew,
T. C. H. Topological polaritons and excitons in garden-variety systems.
\textit{Phys. Rev. B} $\textbf{91}$, 161413(R) (2015).

\bibitem{Nalitov15} Nalitov, A. V., Solnyshkov, D. D. \& Malpuech,
G. Polariton $\mathbb{Z}$ Topological Insulator \textit{Phys. Rev.
Lett.} $\textbf{114}$, 116401 (2015).

\bibitem{Ryu02} Ryu, S. \& Hatsugai, Y. Topological Origin of Zero-Energy
Edge States in Particle-Hole Symmetric Systems. \textit{Phys. Rev.
Lett.} $\textbf{89}$, 077002 (2002).

\bibitem{Li12} Li, H., Peng, H.-L. \& Liu, Z.-F. Two-Dimensional
Nanostructures of Topological Insulators and Their Devices. \textit{Acta
Physico-Chemica Sinca} $\textbf{28}$, 2423 (2012).

\bibitem{Bery06} Brey, L. \& Fertig, H. A. Electronic states of graphene
nanoribbons studied with the Dirac equation. \textit{Phys. Rev. B}$\textbf{73}$,
235411 (2006).

\bibitem{Peres06} Peres, N. M. R., Castro Neto, A. H. \& Guinea,
F. Conductance quantization in mesoscopic graphene. \textit{Phys.
Rev. B} $\textbf{73}$, 195411 (2006); Electronic properties of disordered
two-dimensional carbon. \textit{Phys. Rev. B} $\textbf{73}$, 125411
(2006).

\bibitem{Neto09} Castro Neto, A. H. et al., The electronic properties of graphene.
\textit{Rev. Mod. Phys.} $\textbf{81}$, 109 (2009).

\bibitem{Castro08} Castro, E. V. et al., Localized States at Zigzag
Edges of Bilayer Graphene. \textit{Phys. Rev. Lett.} $\textbf{100}$,
026802 (2008).

\bibitem{Yao09} Yao, W., Yang, S. A. \& Niu, Q. Edge States in Graphene:
From Gapped Flat-Band to Gapless Chiral Modes. \textit{Phys. Rev.
Lett.} $\textbf{102}$, 096801 (2009).

\bibitem{Qiao12} Qiao, Z., Tse, W.-K., Jiang, H., Yao, Y. \& Niu,
Q. Two-Dimensional Topological Insulator State and Topological Phase
Transition in Bilayer Graphene. \textit{Phys. Rev. Lett.} $\textbf{107}$,
256801(2011).

\bibitem{Xiao12} Xiao, D. et al.,
Coupled Spin and Valley Physics in Monolayers of MoS2 and Other Group-VI
Dichalcogenides. \textit{Phys. Rev. Lett.} $\textbf{108}$, 196802
(2012).

\bibitem{Xu14} Xu, X., Yao, W., Xiao, D. \& Heinz, T. F. Spin and
pseudospins in layered transition metal dichalcogenides. \textit{Nat.
Phys.} $\textbf{10}$, 343 (2014).

\bibitem{Liu15} Liu, G.-B. et al., Electronic
structures and theoretical modelling of two-dimensional group-VIB
transition metal dichalcogenides. \textit{Chem. Soc. Rev.}$\textbf{44}$,
2643 (2015).

\bibitem{Mak12} Mak, K. F., He, K., Shan, J. \& Heinz, T. F. Control
of valley polarization in monolayer $\mathrm{MoS_{2}}$ by optical
helicity \textit{Nat. Nanotechnol.}$\textbf{7}$, 494-498 (2012).

\bibitem{Zeng12} Zeng, H. et al.,
Valley polarization in $\mathrm{MoS_{2}}$ monolayers by optical pumping.
\textit{Nat. Nanotechnol.} $\textbf{7}$, 490 (2012).

\bibitem{Cao12}Cao, T. et al., J. Valley-selective
circular dichroism of monolayer molybdenum disulphide. \textit{Nat.
Commum.} $\textbf{3}$, 887 (2012).

\bibitem{Jones13} Jones, A. M. et al., Optical generation of excitonic valley coherence in monolayer
$\mathrm{WSe_{2}}$. \textit{Nat. Nanotechnol.} $\textbf{8}$, 634-638
(2013).

\bibitem{Wang14} Wang, G. et al., Valley dynamics probed through charged and neutral exciton
emission in monolayer $\mathrm{WSe_{2}}$. \textit{Phys. Rev. B} \textbf{90}, 075413 (2014).

\bibitem{Kadantsev12} Kadantsev, E. S. \& Hawrylakb, P., Electronic structure of a single MoS2 monolayer. \textit{Solid State Comm.} \textbf{152}, 909 (2012).

\bibitem{Niu15} Qiu, D. Y., Cao, T. \& Louie, S. G., Nonanalyticity, Valley Quantum Phases, and Lightlike Exciton Dispersion in Monolayer Transition Metal Dichalcogenides: Theory and First-Principles Calculations. \textit{Phys. Rev. Lett.} \textbf{115}, 176801 (2015).

\bibitem{Aivazian15} Aivazian, G. et al., Magnetic control of valley pseudospin in monolayer $\mathrm{WSe_{2}}$.
\textit{Nat. Phys.} $\textbf{11}$,148 (2015).

\bibitem{Yu15}Yu H., Cui X., Xu X., \& Yao W., Valley excitons in two-dimensional semiconductors. \textit{National Science Review} $\textbf{2}$, 57 (2015).

\bibitem{Yu14} Yu, H. et al., Dirac
cones and Dirac saddle points of bright excitons in monolayer transition
metal dichalcogenides. \textit{Nat. Commum}. $\textbf{5}$, 3876 (2014).

\end{thebibliography}
\end{document}